\documentclass[%
 reprint,
superscriptaddress,
 amsmath,amssymb,
 aps,
pra,
floatfix,
]{revtex4-1}

\usepackage{graphicx}
\usepackage{dcolumn}
\usepackage{hyperref}
\usepackage{bm}

\begin{document}

\preprint{APS/123-QED}

\title{Superradiance of molecular nitrogen ions in strong laser fields}

\author{Quanjun Wang}
\thanks{wangqj15@lzu.edu.cn}
\affiliation{EP Department, CERN, CH-1211 Geneva 23, Switzerland}
\affiliation{School of Nuclear Science and Technology, Lanzhou University, Lanzhou 730000, People's Republic of China}
\author{Pengji Ding}
\thanks{dingpj@lzu.edu.cn}
\affiliation{School of Nuclear Science and Technology, Lanzhou University, Lanzhou 730000, People's Republic of China}
\author{Shane G. Wilkins}
\affiliation{Department of Physics, Massachusetts Institute of Technology, Cambridge, MA 02139, USA}
\author{Michail Athanasakis-Kaklamanakis}
\affiliation{EP Department, CERN, CH-1211 Geneva 23, Switzerland}
\affiliation{KU Leuven, Instituut voor Kern- en Stralingsfysica, B-3001 Leuven, Belgium}
\author{Yuxuan Zhang}
\author{Zuoye Liu}
\author{Bitao Hu}
\affiliation{School of Nuclear Science and Technology, Lanzhou University, Lanzhou 730000, People's Republic of China}

\date{\today}

\begin{abstract}

We perform a combined theoretical and experimental investigation of the superradiance in the quantum coherent system generated by strong laser fields. The semiclassical theory of superradiance that includes the superradiant temporal profile, character duration, time delay, intensity is derived. The experimental data and theoretical predictions of 391-nm forward emission as a function of nitrogen gas pressure are compared and show good agreement. Our results not only demonstrate that the time-delayed optical amplification inside the molecular nitrogen ions is superradiance, but also reveal the quantum optical properties of strong-field physics.

\end{abstract}

\pacs{Valid PACS appear here}
\maketitle

Over the past decade, numerous articles reported and/or discussed the 391-nm ``lasing'' action of molecular nitrogen ions in strong femtosecond laser fields~\cite{luo2003lasing,yao2011high,liu2013self,yao2013remote,li2014signature,xu2015sub,liu2015recollision,yao2016population,azarm2017optical,britton2018testing,mysyrowicz2019lasing,tikhonchuk2020theory,chen2021electronic}. The forward emission inside the underdense plasma of pure nitrogen, experiences an increase of energy by several orders of magnitude compared to the seed pulse at 391\,nm~\cite{liu2015recollision,yao2016population}. The time-resolved measurements show that the seed pulse is almost unaffected after passing through the plasma, but it triggers a retarded emission instead~\cite{li2014signature,liu2015recollision}. This emission following the seed pulse has some notable features. After the seed pulse, the emitted intensity increases gradually and reaches its peak at a time delay $\tau_D$ of several picoseconds~\cite{li2014signature,liu2015recollision}. The duration $\tau_W$ of emission shares the same magnitude as $\tau_D$. The experiments at low pressures show that $\tau_W$ is inversely proportional to the plasma length, and the peak intensity scales like the square of the nitrogen gas pressure~\cite{li2014signature}.

These results indicate that the 391-nm ``lasing'' behaves like the Dicke superradiance, which describes the  cooperative emission of photons from a collection of molecules~\cite{dicke1954coherence}. Regarding the superradiance, the emitted power and intensity scales as $N^2$ and the duration is proportional to $N^{-1}$ with $N$ denoting the number density of emitters, which is a coherent radiation. As discussed by Robert Dicke, a cooperation number $r$ was introduced to characterize the coherence of correlation. The cooperation number is completely a quantum effect and is integral or half-integral. Choosing the energy eigenvalues of upper and lower states as $\frac{1}{2}\hbar\omega$ and $-\frac{1}{2}\hbar\omega$ with $\hbar$ and $\omega$ being the Plank constant and transition frequency, $\mid m\mid\leq r\leq\frac{1}{2}N$ is obtained~\cite{dicke1954coherence}, where $m$ is the energy of $N$ molecules in units of $\hbar\omega$. For a system with a certain cooperation number $r$, the superradiance can be studied by investigating the energy of system. A good approximation is the semi-classical theory below with $r$ approaching $\frac{N}{2}$.

It is noticed that the transverse relaxation (dephasing) and the longitidinal relaxation including spontaneous emission and nonradiative decay both can weaken or even eliminate the emission of superradiance~\cite{malcuit1987transition,bonifacio1971quantum,macgillivray1976theory,polder1979superfluorescence,bonifacio1975cooperative}. The dephasing time is much smaller than that of longitidinal relaxation. The radiation is characteristic of superradiance and/or superfluorescence if dephasing time $\gg \sqrt{\tau_W \tau_D}$~\cite{malcuit1987transition,schuurmans1979superfluorescence,schuurmans1980superfluorescence}. The dephasing is mainly caused by electron-ion collision in the nitrogen plasma generated by femtosecond laser pulses. The ponderomotive potential of electron is $\sim$6 eV ($\sim10^8~\mathrm{cm/s}$) in a linearly polarized laser field with the intensity of $10^{14}~\mathrm{W/cm^2}$~\cite{chin2010femtosecond,mitryukovskiy2015plasma}. The collision cross section between nitrogen molecules and free electrons is $\sim10^{-15}~\mathrm{cm^2}$~\cite{itikawa2006cross}. By assuming that the collision cross section $\sigma$ between molecular nitrogen ions and free electrons is the same value, the mean time between collisions is $\frac{1}{\sigma N_i v_e}$, where $N_i$ and $v_e$ are the density of molecular ions and the free electron velocity. It is $\sim$ 200 ps for the nitrogen pressure of 20 mbar with 10\% gas ionization~\cite{mysyrowicz2019lasing}, which is the lower limit of the dephasing time. In the theoretical treatment, the relaxation time far larger than $\tau_D$ and $\tau_W$ is believed and neglected.

The interaction between the seed pulse and the two-level system of $\mathrm{N_2^+}(B^2\Sigma_{u}^+,\nu'=0)$ and $\mathrm{N_2^+}(X^2\Sigma_{g}^+,\nu=0)$ can be expressed by optical Bloch equations~\cite{allen1987optical,cohen1998atom}. The evolution of the system is described by the Bloch angle
\begin{equation}
    \theta(t)=\frac{\mu E_0}{\hbar}\int_0^t f(t') dt'=\int_0^t \Omega(t') dt'
    \label{Bloch_angle}, 
\end{equation}
where $\mu$, $E_0$ and $f(t)$ denote the transition dipole matrix element, the peak and the envelope of the electric field of seed pulse, respectively. After interactions with the seed laser, the system has a Bloch angle $\theta(\tau_r)$, where $\tau_r$ is the interaction time. The Bloch angle does not disappear immediately but develops with time. we obtain the evolution ($t>\tau_r$) of the Bloch angle as well as the two level system by considering a pencil-shaped geometry for the active volume, i.e., $r\ll L$, where $r$ and $L$ are the radius and length of the plasma. It is reasonable because the plasma radius is usually less than 100 $\mu$m, and $L$ is in the millimeter scope with short-focus lenses~\cite{wang2021populations}. The detailed derivations and solutions can be found in the Supplemental Material~\cite{SupplementalMaterial}. 

The Bloch angle is 
\begin{equation}
    \theta(t)=2\mathrm{arctan}(e^{\frac{t-\tau_D}{\tau_W}})
    \label{Bloch_angle solved},
\end{equation}
where 
\begin{equation}
    \tau_W=\frac{4\hbar}{\mu_0 c {\omega}\mu^2 w_0 N L}
    \label{tau_W}
\end{equation}
is the characteristic duration of superradiance with $\mu_0$ and $c$ denoting the vacuum permeability and the speed of light. $\omega$ and $w_0$ are the transition frequency and the initial population probability difference between $\mathrm{N_2^+}(B^2\Sigma_{u}^+,\nu'=0)$ and $\mathrm{N_2^+}(X^2\Sigma_{g}^+,\nu=0)$, respectively. $N$ is the sum of population density of $\mathrm{N_2^+}(B^2\Sigma_{u}^+,\nu'=0)$ and $\mathrm{N_2^+}(X^2\Sigma_{g}^+,\nu=0)$. The characteristic duration can be expressed by the spontaneous decay time $\tau_{sp}$
\begin{equation}
    \tau_W=\frac{16\pi \tau_{sp}}{3\lambda^2 w_0 N L},
\end{equation}
where $\tau_{sp}$ is $\frac{3\pi \varepsilon_0 \hbar c^3}{\omega^3 \mu^2}$ with $\varepsilon_0$ representing the vacuum permittivity~\cite{van2005frequency}. The time delay $\tau_D$ of superradiance is 
\begin{eqnarray}
    \tau_D&&=\tau_r-\mathrm{ln}[\mathrm{tan}\frac{\theta(\tau_r)}{2}]\tau_W \nonumber\\
    &&=\tau_r- \frac{4\hbar \mathrm{ln}[\mathrm{tan}\frac{\theta(\tau_r)}{2}]}{\mu_0 c {\omega}\mu^2 w_0 N L}
    \label{tau_D solved}.
\end{eqnarray}
The energy density of the two-level is
\begin{equation}
    E_N(t)=-\frac{1}{2}\hbar \omega w_0 N \mathrm{tanh}(\frac{t-\tau_D}{\tau_W}).
    \label{E_N_t}
\end{equation}
The power $P_s$ of superradiance per unit volume is
\begin{equation}
    P_s=\frac{1}{8}{\mu_0 c {\omega}^2\mu^2 {w_0}^2 N^2 L}\mathrm{sech}^2(\frac{t-\tau_D}{\tau_W})
    \label{P_s}.
\end{equation}
The intensity $I_s$ of superradiance is 
\begin{equation}
    I_s=\frac{1}{8}{\mu_0 c {\omega}^2\mu^2 {w_0}^2 N^2 L^2}\mathrm{sech}^2(\frac{t-\tau_D}{\tau_W})
    \label{I_s}.
\end{equation}

\begin{figure*}[ht]
\centering
\includegraphics[width=\textwidth]{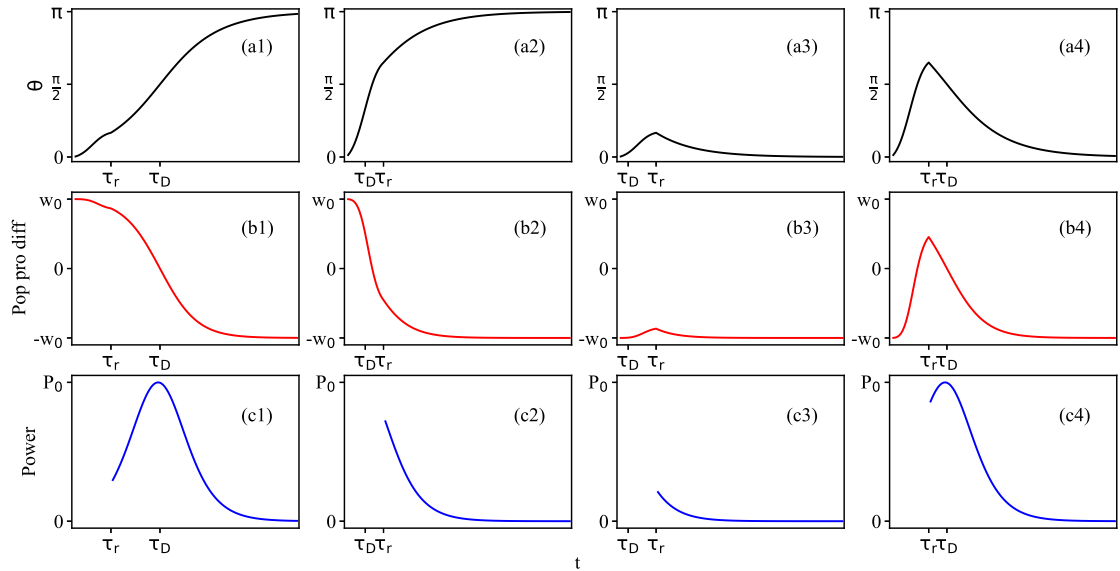}
\caption{Evolutions of (a) the Bloch angle, (b) the population probability difference and (c) the emitted power per volume of superradiance for (1) $w_0>0$, $\theta(\tau_r)<\frac{\pi}{2}$, (2) $w_0>0$, $\theta(\tau_r)>\frac{\pi}{2}$, (3) $w_0<0$, $\theta(\tau_r)<\frac{\pi}{2}$ and (4) $w_0<0$, $\theta(\tau_r)>\frac{\pi}{2}$. The seed pulse is taken as Gaussian beam shape; the envelope electric field is $f(t)=e^{-2\mathrm{ln}2(\frac{t-\tau_s}{\tau_s})^2}$, where $\tau_s$ is the full width at half maximum (FWHM) of spectral intensity. The interaction time is set to be $\tau_r=2\tau_s$. The peak power of superradiance is $P_0=\frac{1}{8}\mu_0 c {\omega}^2\mu^2 {w_0}^2 N^2 L$.}
\label{energy-power}
\end{figure*}

The time-delayed emission can be understood as follows. As described by Eq.~(\ref{E_N_t}), there are still energy stored in the system after interaction with the seed pulse. The release of remaining energy results in the retarded emission. It can be seen that the initial population probability difference $w_0$, determined by the pump pulse, and Bloch angle $\theta(\tau_r)$, caused by the seed pulse, govern the temporal evolution of the energy as well as the superradiance. According to the source of energy, the superradiance is discussed in two cases of $w_0>0$ and $w_0<0$.

For $w_0>0$, if the seed pulse is weak, the Bloch angle $\theta(\tau_r)$ is less than $\frac{\pi}{2}$, the time delay $\tau_D$ is larger than $\tau_r$, and the Bloch angle goes to $\pi$ in Fig.~\ref{energy-power}(a1). The population probability difference $w$ decreases from $w_0$ to $-w_0$ as displayed in Fig.~\ref{energy-power}(b1). Accompanied by the seed pulse, only part energy $\frac{1}{2}\hbar \omega w_0 N  [1-\mathrm{cos}\theta(\tau_r)]$ is emitted. The system radiates the remaining energy based on Eqs.~(\ref{P_s}) and/or (\ref{I_s}), which is the origination of 391-nm optical amplification inside molecular nitrogen ions. Figure~\ref{energy-power}(c1) shows the power of superradiance, whose peak power $P_0$ appears at the time delay of $\tau_D$. As the power is proportional to the square of the electric field of superradiance, figure~\ref{energy-power}(c1) indicates the development of the macro dipole inside the system. It achieves the maximal value at $\tau_D$, where the population of $\mathrm{N_2^+}(B^2\Sigma_{u}^+,\nu'=0)$ equals that of $\mathrm{N_2^+}(X^2\Sigma_{g}^+,\nu=0)$, the Bloch angle increases fastest and the energy is released most quickly. If the seed pulse is strong, the Bloch angle $\theta(\tau_r)$ is larger than $\frac{\pi}{2}$, the time delay $\tau_D$ is less than $\tau_r$, and the Bloch angle still goes to $\pi$ in Fig.~\ref{energy-power}(a2). There are a lot of excited-state decays with the presence of seed pulse, which amplifies the seed pulse efficiently in Fig.~\ref{energy-power}(b2). The power could not reach $P_0$ but declines from a certain value as illustrated in Fig.~\ref{energy-power}(c2). For $w_0>0$, the seed pulse plays a role of trigger, triggering the release of energy of $N w_0 \hbar \omega$.
 
For $w_0<0$, the Bloch angle $\theta(\tau_r)$ and the time delay $\tau_D$ are, respectively, smaller than $\frac{\pi}{2}$ and $\tau_r$ with the weak seed pulse. The Bloch angle, the population probability difference and the superradiant power all decrease in Fig.~\ref{energy-power}(a3)--(c3). With the strong seed pulse, the Bloch angle $\theta(\tau_r)$ and the time delay $\tau_D$ are larger than $\frac{\pi}{2}$ and $\tau_r$, respectively [Fig.~\ref{energy-power}(a4)]. The $w$ drops from $w(\tau_r)>0$ [Fig.~\ref{energy-power}(b4)] and the power gets to the peak $P_0$ at $\tau_D$ [Fig.~\ref{energy-power}(c4)]. Unlike the case of $w_0>0$, the energy of emissions for $w_0<0$ totally comes from the seed pulse. The two-level system acts like a battery---rapid storage and slow release of energy. 

\begin{table*}[ht]
\caption{The experimental FWHM $\tau_{FW}$ and time delay $\tau_D$ of the 391-nm forward emission as a function of nitrogen pressure $p$.}
\begin{ruledtabular}
\begin{tabular}{cccccccccc}
 $p$ (mbar) &6 &7 &8 &10 &12 &14 &16 &18 &20\\ \hline
 $\tau_{FW}$ (ps)  &3.995 &3.508 &2.937 &2.349 &2.001 &1.581 &1.303 &1.082 &1.003\\ \hline  
 $\tau_D$ (ps)  &8.614 &7.295 &6.287 &4.613 &3.822 &3.070 &2.701 &2.450 &2.199\\
\end{tabular}
\end{ruledtabular}
\label{tab}
\end{table*}

\begin{figure*}[ht]
\centering
\includegraphics[width=\textwidth]{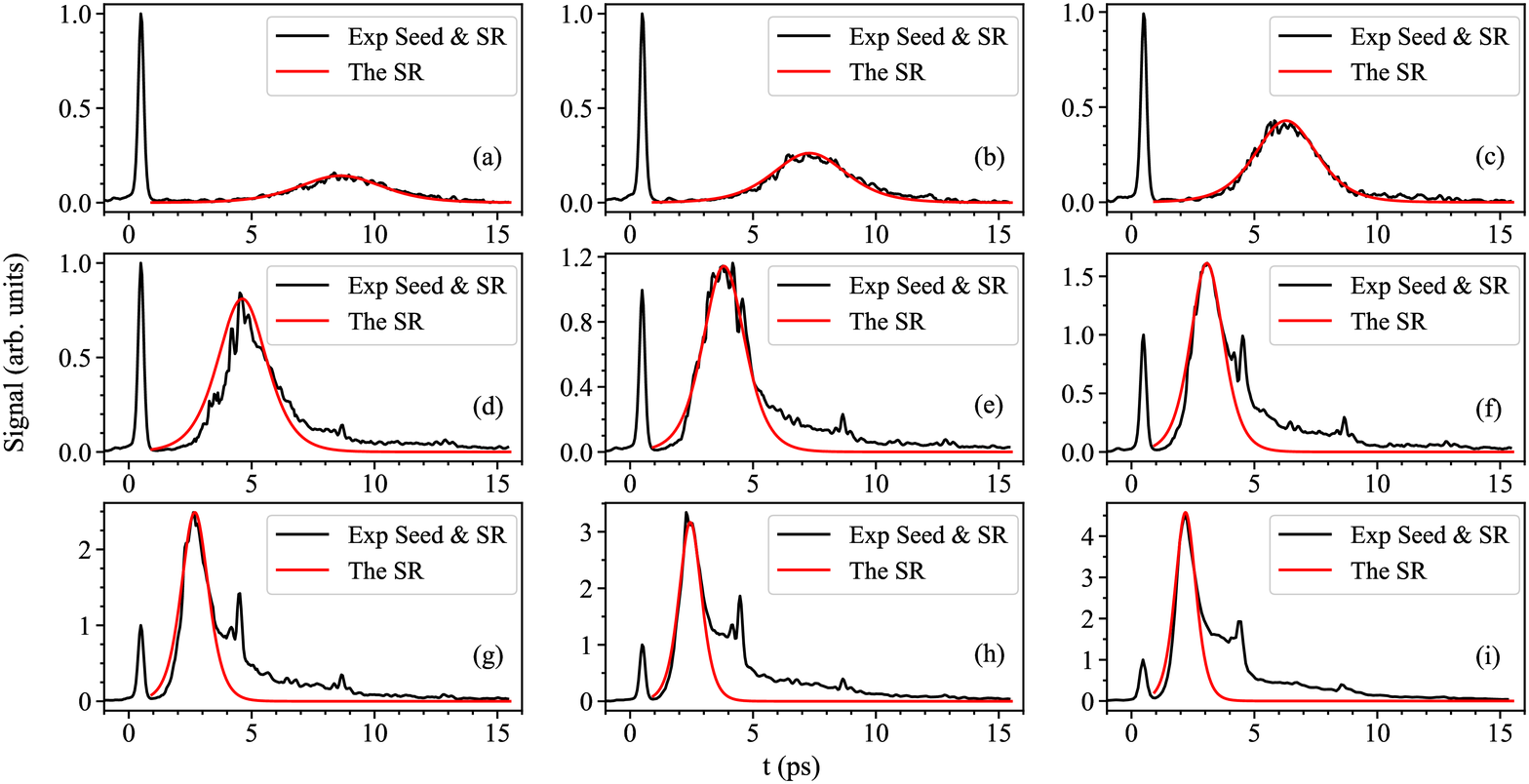}
\caption{Comparisons of the experimental data and theoretical predictions for the temporal profile of the 391-nm forward emission with different gas pressure of (a) 6 mbar, (b) 7 mbar, (c) 8 mbar, (d) 10 mbar, (e) 12 mbar, (f) 14 mbar, (g) 16 mbar, (h) 18 mbar and (i) 20 mbar, respectively.}
\label{temporal-pre}
\end{figure*}

Next, we perform comparisons of the experimental data and theoretical prediction for the temporal profile of the 391-nm forward emission. The experiment investigated the time-resolved signals as a function of nitrogen pressure from 6 to 20 mbar in Fig.~\ref{temporal-pre}. The details of the experiment was described in Ref~\cite{ding2016lasing}. The seed pulse with the FWHM $\tau_s=0.26$ ps can be expressed by a Gaussian profile. The interaction time $\tau_r$ between the seed pulse and the two-level system is 3.6$\tau_s$. Table~\ref{tab} lists the experimental FWHM $\tau_{FW}$ and time delay $\tau_D$ of the 391-nm forward emission at different pressures. The character duration $\tau_W=\frac{\tau_{FW}}{1.763}$ is used for the hyperbolic secant pulse. By substituting $\tau_W$ and $\tau_D$ into Eq.~(\ref{I_s}), we obtain the temporal profile of superradiance. As shown in Fig.~\ref{temporal-pre}(a)--(c), the theoretical results agree well with the experimental data at pressures of 6--8 mbar, which is a strong evidence that the 391-nm emission inside ionized nitrogen molecules is superradiance. When the gas pressure exceeds 10 mbar, the theoretical predictions are in agreement with the experimental signals for the main part of superradiance, as illustrated in Fig.~\ref{temporal-pre}(d)--(i). Following the strongest radiation, there are the other two gains at 4.2 and 8.4 ps. This behaviour is caused by modulation of rotational coherence~\cite{zhang2013rotational,zhang2019coherent}.

\begin{figure}[ht]
\centering
\includegraphics[width=3.375in]{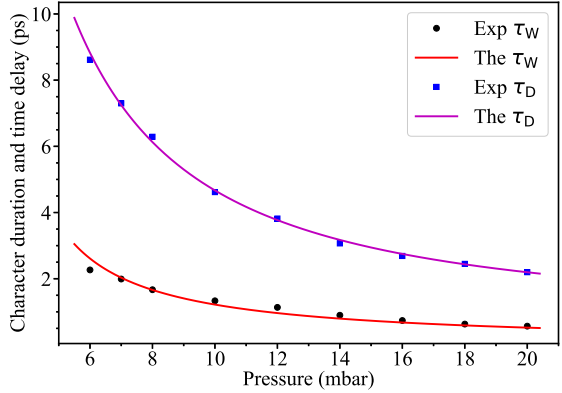}
\caption{Comparisons between the experimental character duration $\tau_W$ (black circle) and time delay $\tau_D$ (blue square), and the calculated $\tau_W$ (red line) and $\tau_D$ (magenta line) as a function of gas pressure.}
\label{Tau_W and tau_D}
\end{figure}

The character duration $\tau_W$ dependent on the gas pressure is calculated according to Eq.~(\ref{tau_W}). The time-dependent quantum-wave-packet calculations show that $w_0$ varies from 0 to 0.3, respectively, corresponding to the 800-nm laser intensity ranging from 2.2 to $4\times 10^{14}$ W/cm$^2$~\cite{xu2015sub}. The pump laser intensity and plasma length $L=10$ mm in our experiment are assumed to be unchanged by using a lens with the focal length of 400 mm~\cite{chin2010femtosecond,mysyrowicz2019lasing}. For simplicity, $w_0$ is considered as a parameter and set to be 0.1. Because of the unchanged laser intensity, the two-level-system population density is proportional to gas pressure. In fact, the superradiance is a collective effect and will vanish when the pressure is lower than a certain value $p_0$. The minimal gas pressure that causes the superradiance is set to be $p_0=2.5$ mbar in the current case. The relationship between the two-level-system population density and pressure is $N=k(p-p_0)$, where $k$ is a scale factor. Using the experimental point ($p=8$ mbar, $\tau_W=1.666$ ps), we obtain $N=0.228\times (p-2.5)(\mathrm{mbar})\times 10^{16}~\mathrm{cm}^{-3}$ with $\mu=1.7$ D~\cite{langhoff1988theoretical}. Therefore, $\tau_W$ (red line) as a function of the nitrogen pressure are computed. The calculated $\tau_W$ (red line) agrees well with the experimental data (black circle), as shown in Fig.~\ref{Tau_W and tau_D}. The character duration is inversely proportional to the population density of the two-level system, a notable feature of superradiance. 

The electric field of the seed pulse can be written as $E(t)=E_0e^{-2\mathrm{ln}2(\frac{t-\tau_s}{\tau_s})^2}$, with the peak of electric field $E_0=\sqrt{\frac{2I_{seed}}{\varepsilon_0 c}}$. The intensity of seed pulse $I_{seed}$ to trigger the 391-nm superradiance is estimated to be 10 MW/\,$\mathrm{cm^2}$. Then the initial Bloch angle is $\theta(\tau_r)=\frac{\mu}{\hbar}\int_0^{\tau_r} E(t) dt=0.057\pi$, far smaller than $\frac{\pi}{2}$; the actual radiation indeed obeys the curve in Fig.~\ref{energy-power}(c1). The time delay $\tau_D$ works out by using Eq.~(\ref{tau_D solved}), as illustrated in Fig.~\ref{Tau_W and tau_D}. It is seen that our theoretical expectations (magenta line) are in good agreement with the experimental results (blue square).

\begin{figure}[ht]
\centering
\includegraphics[width=3.375in]{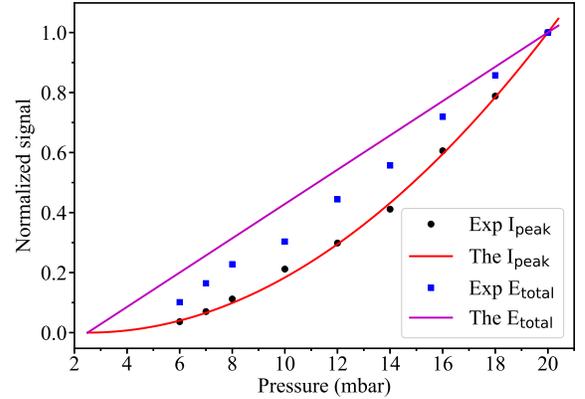}
\caption{Comparisons between the experimental results of peak intensity $I_{peak}$ (black circle) and total emitted energy $E_{total}$ (blue square), and the theoretical calculations of $I_{peak}$ (red line) and $E_{total}$ (magenta line) as a function of gas pressure.}
\label{I_p and E_t}
\end{figure}

The peak intensity $I_{peak}$ and total emitted energy $E_{total}$ of the superradiance are investigated as a function of nitrogen pressure (or the population density $N$ of the two-level system). The values at $t=\tau_D$ and the integrals of the 391-nm emission in Fig.~\ref{temporal-pre} are taken as the peak intensity and the total emitted energy of the experiment, respectively. They are normalized such that their maximal values at 20 mbar equal 1, as shown in Fig.~\ref{I_p and E_t}. The peak intensity is expressed as $I_{peak}=\frac{1}{8}{\mu_0 c {\omega}^2 \mu^2 {w_0}^2 N^2 L^2}$, which is proportional to $N^2$. The total emitted energy $E_{total}$ is $\hbar \omega N w_0 \mathrm{cos}\theta({\tau_r})$ multiplied by the active volume, which is proportional to $N$. Using the linear relationship between $N$ and $p$, the normalized $I_{peak}$ (read line) and $E_{total}$ (magenta line) of the theory are calculated, as shown in Fig.~\ref{I_p and E_t}. The good agreement of the experimental data and calculated results confirms the superradiance nature of the 391-nm forward emission.

In conclusion, we theoretically investigate the evolution of energy in the coherent system of molecular nitrogen ions through solving the Bloch angle. The semiclassical superradiance theory is proposed to describe the superradiant character duration, time delay and intensity. We explain the time-delayed optical amplification in molecular nitrogen ions irradiated with intense femtosecond laser pulses, and reveal the superradiance nature of the 391-nm forward emission by the comparisons between the theoretical and experimental results. Our findings provide direct evidences of molecular coherence in strong laser field.

This work was supported by China Scholarship Council, and the National Natural Science Foundation of China (Grants No. U1932133, No. 11905089, and No. 12004147).


\begin{thebibliography}{99}
\bibitem{luo2003lasing} Q. Luo, W. Liu, and S. L. Chin, Appl. Phys. B \textbf{76}, 337 (2003).

\bibitem{yao2011high} J. Yao, B. Zeng, H. Xu, G. Li, W. Chu, J. Ni, H. Zhang, S. L. Chin, Y. Cheng, and Z. Xu, Phys. Rev. A \textbf{84}, 051802(R) (2011).

\bibitem{liu2013self} Y. Liu, Y. Brelet, G. Point, A. Houard, and A. Mysyrowicz, Opt. Express \textbf{21}, 22791 (2013).

\bibitem{yao2013remote} J. Yao, G. Li, C. Jing, B. Zeng, W. Chu, J. Ni, H. Zhang, H. Xie, C. Zhang, H. Li, H. Xu, S. L. Chin, Y. Cheng, and Z. Xu, New J. Phys. \textbf{15}, 023046 (2013).

\bibitem{li2014signature} G. Li, C. Jing, B. Zeng, H. Xie, J. Yao, W. Chu, J. Ni, H. Zhang, H. Xu, Y. Cheng, and Z. Xu, Phys. Rev. A \textbf{89}, 033833 (2014).

\bibitem{xu2015sub} H. Xu, E. L{\"o}tstedt, A. Iwasaki, and K. Yamanouchi, Nat. Commun. \textbf{6}, 8347 (2015).

\bibitem{liu2015recollision} Y. Liu, P. Ding, G. Lambert, A. Houard, V. Tikhonchuk, and A. Mysyrowicz, Phys. Rev. Lett. \textbf{115}, 133203 (2015).

\bibitem{yao2016population} J. Yao, S. Jiang, W. Chu, B. Zeng, C. Wu, R. Lu, Z. Li, H. Xie, G. Li, C. Yu, Z. Wang, H. Jiang, Q. Gong, and Y. Cheng, Phys. Rev. Lett. \textbf{116}, 143007 (2016).

\bibitem{azarm2017optical} A. Azarm, P. Corkum, and P. Polynkin, Phys. Rev. A \textbf{96}, 051401(R) (2017).

\bibitem{britton2018testing} M. Britton, P. Laferriere, D. H. Ko, Z. Li, F. Kong, G. Brown, A. Naumov, C. Zhang,  L. Arissian, and P. B. Corkum, Phys. Rev. Lett. \textbf{120}, 133208 (2018).

\bibitem{mysyrowicz2019lasing} A. Mysyrowicz, R. Danylo, A. Houard, V. Tikhonchuk, X. Zhang, Z. Fan, Q. Liang, S. Zhuang, L. Yuan, and Y. Liu, APL Photonics \textbf{4}, 110807 (2019).

\bibitem{tikhonchuk2020theory} V. T. Tikhonchuk, Y. Liu, R. Danylo, A. Houard, and A. Mysyrowicz, arXiv:2003. 12840 (2020).

\bibitem{chen2021electronic} J. Chen, J. Yao, Z. Zhang, Z. Liu, B. Xu, Y. Wan, F. Zhang, W. Chu, L. Qiao, H. Zhang, Z. Wang, and Y. Cheng, Phys. Rev. A \textbf{103}, 033105 (2021).

\bibitem{dicke1954coherence} R. H. Dicke, Phys. Rev. \textbf{93}, 99 (1954).

\bibitem{malcuit1987transition} M. S. Malcuit, J. J. Maki, D. J. Simkin, and R. W. Boyd, Phys. Rev. Lett. \textbf{59}, 1189 (1987).

\bibitem{bonifacio1971quantum} R. Bonifacio, P. Schwendimann, and F. Haake, Phys. Rev. A \textbf{4}, 302 (1971).

\bibitem{macgillivray1976theory} J. C. MacGillivray and M. Feld, Phys. Rev. A \textbf{14}, 1169 (1976).

\bibitem{polder1979superfluorescence} D. Polder, M. F. H. Schuurmans, and Q. H. F. Vrehen, Phys. Rev. A \textbf{19}, 1192 (1979).

\bibitem{bonifacio1975cooperative} R. Bonifacio and L. Lugiato, Phys. Rev. A \textbf{11}, 1507 (1975).

\bibitem{schuurmans1979superfluorescence} M. F. H. Schuurmans and D. Polder, Phys. Lett. A \textbf{72}, 306 (1979).

\bibitem{schuurmans1980superfluorescence} M. F. H. Schuurmans, Opt. Commun. \textbf{34}, 185 (1980).

\bibitem{chin2010femtosecond} S. L. Chin, Femtosecond laser filamentation, Vol. 55 (Springer, 2010).

\bibitem{mitryukovskiy2015plasma} S. Mitryukovskiy, Y. Liu, P. Ding, A. Houard, A. Couairon, and A. Mysyrowicz, Phys. Rev. Lett. \textbf{114}, 063003 (2015).

\bibitem{itikawa2006cross} Y. Itikawa, J. Phys. Chem. Ref. Data \textbf{35}, 31 (2006).

\bibitem{allen1987optical} L. Allen and J. H. Eberly, \textit{Optical resonance and two-level atoms} (Dover, New York, 1987).

\bibitem{cohen1998atom} C. Cohen-Tannoudji, J. Dupont-Roc, and G. Grynberg, \textit{Atom-photon interactions: basic processes and applications} (Wiley, New York, 1998).

\bibitem{wang2021populations} Q. Wang, R. Chen, Y. Zhang, X. Wang, C. Sun, P. Ding, Z. Liu, and B. Hu, Phys. Rev. A \textbf{103}, 033117 (2021).
 
\bibitem{SupplementalMaterial} See Supplemental Materialatht at http://link.aps.org/supplemental/XXXX for the details of theoretical derivation.
 
\bibitem{van2005frequency} A. F. van Driel, G. Allan, C. Delerue, P. Lodahl, W. L. Vos, and D. Vanmaekelbergh, Phys. Rev. Lett. \textbf{95}, 236804 (2005).

\bibitem{ding2016lasing} P. Ding, \textit{Lasing  effect  in  femtosecond  filaments  in  air}, Ph.D. thesis, Universit{\'e} Paris-Saclay (2016).

\bibitem{zhang2013rotational} H. Zhang, C. Jing, J. Yao, G. Li, B. Zeng, W. Chu, J. Ni, H. Xie, H. Xu, S. L. Chin, K. Yamanouchi, Y. Cheng, and Z. Xu, Phys. Rev. X \textbf{3}, 041009 (2013).

\bibitem{zhang2019coherent} A. Zhang, Q. Liang, M. Lei, L. Yuan, Y. Liu, Z. Fan, X. Zhang, S. Zhuang, C. Wu, Q. Gong and H. Jiang, Opt. Express \textbf{27}, 12638 (2019).

\bibitem{langhoff1988theoretical} S. R. Langhoff and C. W. Bauschlicher Jr, J. Chem. Phys. \textbf{88}, 329 (1988).

\end{thebibliography}
\end{document}


\title{\textbf{Supplementary Material}}


\maketitle

\section{Semiclassical theory of superradiance}
The time-dependent Hamiltonian for describing the laser-induced coherence between the ground and excited electronic states of molecular nitrogen ions is given by
\begin{eqnarray}
H(t)=
\begin{pmatrix}
\frac{1}{2}\hbar\omega  &-\mu E(t)\\
-\mu E(t) & -\frac{1}{2}\hbar\omega
\end{pmatrix}
\label{Hamiltonian},
\end{eqnarray}
where $\frac{1}{2}\hbar\omega$ and $-\frac{1}{2}\hbar\omega$ represent the energy eigenvalues of $\mathrm{N_2^+}(B^2\Sigma_{u}^+,\nu'=0)$ and $\mathrm{N_2^+}(X^2\Sigma_{g}^+,\nu=0)$, respectively. $\mu$ is the transition dipole matrix element and $E(t)$ denotes the seed-laser electric field. The density matrix of the two-level system is defined as
\begin{eqnarray}
\rho(t)=
\begin{pmatrix}
\rho_{22}(t) & \rho_{21}(t)\\
\rho_{12}(t) & \rho_{11}(t)
\end{pmatrix}
\label{Density_matrix}.
\end{eqnarray}
The diagonal elements $\rho_{22}(t)$ and $\rho_{11}(t)$ are probabilities of being in the upper and lower states. The meaning of off-diagonal elements can be revealed by calculating the expectation value of the electric dipole operator. $\rho_{21}(t)+\rho_{12}(t)$ is the electronic transition moment in units of $\mu$. The density operator equation reads 
\begin{equation}
    \frac{\mathrm{d} \rho(t)}{\mathrm{d} t}=-\frac{i}{\hbar}[H(t),\rho(t)]
    \label{Density_matrix_equation}.
\end{equation}
The electric field $E(t)$ of the seed-laser is written as
\begin{equation}
    E(t)=\frac{\bar{E}(t)}{2}(e^{-i\omega t}+e^{i\omega t})
    \label{electric field},
\end{equation}
where $\bar{E}(t)$ is the amplitude of the electric field and it is a real number here. In such laser fields, the electric dipole oscillates at frequency of $\omega$. $\rho_{21}(t)$ and $\rho_{12}(t)$ can be expressed as
\begin{equation}
    \rho_{21}(t)=\bar{\rho}_{21}(t)e^{-i\omega t}
    \label{rho21}
\end{equation}
and 
\begin{equation}
    \rho_{12}(t)=\bar{\rho}_{12}(t)e^{i\omega t}
    \label{rho12},
\end{equation}
with $\bar{\rho}_{21}(t)$ and $\bar{\rho}_{12}(t)$ denoting the amplitudes of $\rho_{21}(t)$ and $\rho_{12}(t)$, respectively.

Under rotating-wave approximation where the high frequency terms $e^{-i2\omega t}$ and $e^{i2\omega t}$can be omitted, we obtain each matrix element of Eq.~(\ref{Density_matrix_equation})
\begin{subequations}
\begin{equation}
\label{partial_rho21}
\frac{\mathrm{d} \bar{\rho}_{21}(t)}{\mathrm{d} t}=-\frac{i\mu \bar{E}(t)}{2\hbar}[\rho_{22}(t)-\rho_{11}(t)],
\end{equation}
\begin{equation}
\label{partial_rho12}
\frac{\mathrm{d} \bar{\rho}_{12}(t)}{\mathrm{d} t}=\frac{i\mu \bar{E}(t)}{2\hbar}[\rho_{22}(t)-\rho_{11}(t)],
\end{equation}
\begin{equation}
\label{partial_rho22}
\frac{\mathrm{d} \rho_{22}(t)}{\mathrm{d} t}=-\frac{i\mu \bar{E}(t)}{2\hbar}[\bar{\rho}_{21}(t)-\bar{\rho}_{12}(t)], 
\end{equation}
\begin{equation}
\label{partial_rho11}
\frac{\mathrm{d} \rho_{11}(t)}{\mathrm{d} t}=\frac{i\mu \bar{E}(t)}{2\hbar}[\bar{\rho}_{21}(t)-\bar{\rho}_{12}(t)]. 
\end{equation}
\end{subequations}
The Bloch components of $u(t)$, $v(t)$ and $w(t)$ are defined as
\begin{subequations}
\begin{equation}
\label{u}
u(t)=\bar{\rho}_{21}(t)+\bar{\rho}_{12}(t),
\end{equation}
\begin{equation}
\label{v}
v(t)=i\bar{\rho}_{21}(t)-i\bar{\rho}_{12}(t),
\end{equation}
\begin{equation}
\label{w}
w(t)=\rho_{22}(t)-\rho_{11}(t). 
\end{equation}
\end{subequations}
It can be seen that $\frac{\mathrm{d} u(t)}{\mathrm{d} t}=\mathrm{Eq}.~(\ref{partial_rho21})+\mathrm{Eq}.~(\ref{partial_rho12})$, $\frac{\mathrm{d} v(t)}{\mathrm{d} t}=i\mathrm{Eq}.~(\ref{partial_rho21})-i\mathrm{Eq}.~(\ref{partial_rho12})$ and $\frac{\mathrm{d} w(t)}{\mathrm{d} t}=\mathrm{Eq}.~(\ref{partial_rho22})-\mathrm{Eq}.~(\ref{partial_rho11})$. We get the optical Bloch equations
\begin{subequations}
\begin{equation}
\label{Bloch1}
\frac{\mathrm{d} u(t)}{\mathrm{d} t}=0,
\end{equation}
\begin{equation}
\label{Bloch2}
\frac{\mathrm{d} v(t)}{\mathrm{d} t}=\Omega w(t),
\end{equation}
\begin{equation}
\label{Bloch3}
\frac{\mathrm{d} w(t)}{\mathrm{d} t}=-\Omega v(t),
\end{equation}
\end{subequations}
where $\Omega(t)=\frac{\mu}{\hbar}\bar{E}(t)$ is Rabi frequency. The amplitude is written as $\bar{E}(t)=E_0 f(t)$ with $E_0$ and $f(t)$ being the peak and the envelope of the seed pulse, respectively. 

After photonization and couplings by 800-nm pump lasers, the initial probability difference between $\mathrm{N_2^+}(B^2\Sigma_{u}^+,\nu'=0)$ and $\mathrm{N_2^+}(X^2\Sigma_{g}^+,\nu=0)$ is $w_0$. The initial $u$ and $v$ are supposed to be 0. To solve equations~(\ref{Bloch1}--\ref{Bloch3}), a Bloch angle is defined as
\begin{equation}
    \theta(t)=\frac{\mu E_0}{\hbar}\int_0^t f(t') dt'=\int_0^t \Omega(t') dt'
    \label{Bloch_angle}.
\end{equation}
Based on $u(0)=0$, we get $u(t)=0$.
As $v^2(t)+w^2(t)$ satisfies
\begin{equation}
    \frac{\mathrm{d}}{\mathrm{d}t}[v^2(t)+w^2(t)]=2\Omega[v(t)w(t)-w(t)v(t)]=0,
\end{equation}
there is 
\begin{equation}
    v^2(t)+w^2(t)={w_0}^2
    \label{conservation}.
\end{equation}
By substituting Eq.~(\ref{conservation}) in Eq.~(\ref{Bloch3}), we obtain 
\begin{equation}
    \frac{\mathrm{d} w(t)}{\mathrm{d} t}=-\Omega \sqrt{{w_0}^2-w^2(t)}
\end{equation}
or
\begin{equation}
    \frac{\mathrm{d} w(t)}{\sqrt{{w_0}^2-w^2(t)}}=-\Omega \mathrm{d} t
    \label{total differential}.
\end{equation}
Integrating Eq.~(\ref{total differential}) from 0 to $t$, we have
\begin{equation}
    -\mathrm{arccos}\frac{w}{w_0} \Biggr|_0^t=-\mathrm{arccos}\frac{w}{w_0}=-\int_0^t \Omega(t') dt',
\end{equation}
and
\begin{equation}
   w(t)=w_0\mathrm{cos} \theta (t)
   \label{w_solved}.
\end{equation}
Using Eq.~(\ref{conservation}), $v(t)$ is
\begin{equation}
    v(t)=\sqrt{{w_0}^2-{w_0}^2 \mathrm{cos}^2 \theta(t)}= w_0 \mathrm{sin} \theta(t),
\end{equation}
where negative sign is abandoned. After interactions with the seed laser, the solutions of optical Bloch equations~(\ref{Bloch1})--(\ref{Bloch3}) are
\begin{subequations}
\begin{equation}
\label{u_tau_r}
u(t)=0,
\end{equation}
\begin{equation}
\label{v_tau_r}
v(t)=w_0 \mathrm{sin}\theta({t}),
\end{equation}
\begin{equation}
\label{w_tau_r}
w(t)=w_0 \mathrm{cos}\theta({t}).
\end{equation}
\end{subequations}

The energy density of the two-level system is
\begin{equation}
    E_N(t)=\frac{1}{2}\hbar \omega N w_0 \mathrm{cos}\theta({t})
    \label{E_taur}.
\end{equation}
 The polarization of the two level system is 
 \begin{eqnarray}
     P(t,z)=&&N\mu [\bar{\rho}_{21}(t) e^{-i\omega t+ikz}+\bar{\rho}_{12}(t) e^{i\omega t-ikz}]\nonumber\\
     &&=\frac{1}{2}[-iN\mu v(t) e^{-i\omega t+ikz}+c.c.]
     \label{P_taur}
  \end{eqnarray}
with $N$ representing the sum of the population densities of $\mathrm{N_2^+}(B^2\Sigma_{u}^+,\nu'=0)$ and $\mathrm{N_2^+}(X^2\Sigma_{g}^+,\nu=0)$. Maxwell equations tell us that there should be an electric field $E_s(t,z)$, which induces this polarization. The relationship between $E_s(t,z)$ and $P(t,z)$ derived from Maxwell equations is 
\begin{equation}
    \frac{{\partial}^2 E_s}{\partial z^2}-\mu_0\varepsilon_0\frac{{\partial}^2 E_s}{\partial t^2}=\mu_0\frac{{\partial}^2 P}{\partial t^2}
    \label{Maxwell_eq},
\end{equation}
where $\mu_0$ and $\varepsilon_0$ denote vacuum permeability and vacuum permittivity, respectivelly; $z$ is the propagation direction of seed laser as well as superradiance. $E_s(t,z)$, the initial electric field of superradiance, can be formally written as 
\begin{equation}
    E_s(t,z)=\frac{1}{2}[\bar{E}(t,z)e^{-i\omega t+ikz}+c.c.]
    \label{E_s},
\end{equation}
where $\bar{E}(t,z)$ is the envelope amplitude of $E_s(t,z)$. Substituting Eqs.~(\ref{P_taur}) and (\ref{E_s}) into Eq.~(\ref{Maxwell_eq}), the complex conjugate terms can be ignored by rotating-wave approximation. Applying slow amplitude approximation,
\begin{subequations}
    \begin{equation}
     \frac{{\partial}^2\bar{E}_s(t,z)}{\partial t^2}\ll \omega \frac{\partial \bar{E}_s(t,z)}{\partial t} \ll {\omega}^2\bar{E}_s(t,z),
    \end{equation}
\begin{equation}
     \frac{{\partial}^2\bar{E}_s(t,z)}{\partial z^2}\ll k \frac{\partial \bar{E}_s(t,z)}{\partial z} \ll k^2\bar{E}_s(t,z),
    \end{equation}
    \begin{equation}
     \frac{{\partial}^2\bar{P}(t,z)}{\partial t^2}\ll \omega \frac{\partial \bar{P}(t,z)}{\partial t} \ll {\omega}^2\bar{P}(t,z),
    \end{equation}
\end{subequations}
which means the relative change of the envelope amplitude of the electric field and polarization within one optical cycle is much less than 2$\pi$, equation~(\ref{Maxwell_eq}) is simplified to
\begin{equation}
    \frac{\partial \bar{E}_s}{\partial z}+\frac{1}{c}\frac{\partial \bar{E}_s}{\partial t}=\frac{\mu_0\omega c \mu N}{2}v
    \label{simplified}.
\end{equation}

A interaction length that causes a single superradiance is defined as $L$. We assume that the intensity of seed pulse is unchanged within the interaction length $L$. The seed pulse causes the same polarization $v$ at each $z$ ($0<z<L$) point. The interaction time between the two-level system and seed pulse at each $z$ ($0<z<L$) point is $\tau_r$ that is related to the pulse duration of seed laser. By making coordinate transformations as
\begin{equation}
   \eta=t-\frac{z}{c}, \xi=z.
\end{equation}
Equation~(\ref{simplified}) is rewritten as
\begin{equation}
    \frac{\partial \bar{E}_s(\eta,\xi)}{\partial \xi}=\frac{\mu_0\omega c \mu N}{2}v(\eta).
\end{equation}
At $z=0$, there is a $v(\tau_r)$ after the interaction of the seed pulse and two-level system; the new coordinates are $\eta=\tau_r$ and $\xi=0$. At $z=L$, the seed pulse first gets to this point with $\delta t=\frac{L}{c}$, and then generates a $v(\delta t+\tau_r)$ after $\tau_r$; the new coordinates are $\eta=\tau_r+\delta t-\frac{L}{c}=\tau_r$ and $\xi=L$. Then we obtain
\begin{equation}
    \bar{E}_s(\tau_r,L)=\frac{\mu_0\omega c \mu N L}{2}v(\tau_r)
    \label{E_s_L_eta_xi},
\end{equation}
with the initial condition $\bar{E}_s(\tau_r,0)=0$. 

In $(x,z)$ coordinate system, equation~(\ref{E_s_L_eta_xi}) is
\begin{equation}
    \bar{E}_s(\tau_r+\delta t,L)=\frac{\mu_0\omega c \mu N L}{2}v(\tau_r)
    \label{E_s_L_t_z}.
\end{equation}
If we shift the zero time from $z=0$ to $z=L$, equation~(\ref{E_s_L_t_z}) is
\begin{equation}
    \bar{E}_s(\tau_r,L)=\frac{\mu_0\omega c \mu N L}{2}v(\tau_r)
    \label{E_s_L_t_z_shift}.
\end{equation}
The energy flux density $S$ of the electric field at ($\tau_r,L$) is
  \begin{eqnarray}
     S=&&\frac{1}{2}\sqrt{\frac{\varepsilon_0}{\mu_0}}{\bar{E}_s}^2(\tau_r,L)=\frac{\mu_0 c {\omega}^2\mu^2 N^2 L^2}{8}v^2(\tau_r)\\\nonumber
     &&=\frac{\mu_0 c {\omega}^2\mu^2 N^2 L^2}{8}{w_0}^2\mathrm{sin}^2\theta(\tau_r)
     \label{S}.
  \end{eqnarray}
The decrease of energy per unit time of the two-level system at $\tau_r$ based on Eq.~(\ref{E_taur}) is
\begin{equation}
    -\frac{\mathrm{d}E_N}{\mathrm{d}t}=\frac{1}{2}\hbar \omega N w_0 \mathrm{sin}\theta \frac{\mathrm{d}\theta}{\mathrm{d}t}\Biggr|_{t=\tau_r}
    \label{-dE_dt}.
\end{equation}
A pencil-shaped geometry for the active volume is considered,i.e.,
\begin{equation}
    r\ll L,
\end{equation}
with $r$ being the radius. It is reasonable in the plasma because the plasma radius is usually less than 100 $\mu$m, and the plasma length is typically in mm range. From the law of conservation of energy, the decrease of energy per unit time from the active volume equals the energy flux from the cross section, which reads
\begin{equation}
    -\frac{\mathrm{d}E_N}{\mathrm{d}t}\pi r^2 L=S\pi r^2,
\end{equation}
and
\begin{equation}
    \frac{1}{2}\hbar \omega N w_0 \mathrm{sin}\theta \frac{\mathrm{d}\theta}{\mathrm{d}t}\Biggr|_{t=\tau_r}=\frac{\mu_0 c {\omega}^2\mu^2 N^2 L^2}{8L}{w_0}^2\mathrm{sin}^2\theta(\tau_r)
    \label{conservation_E}.
\end{equation}
The derivative of Bloch angle with respect to time is
\begin{equation}
    \frac{\mathrm{d}\theta}{\mathrm{d}t}\Biggr|_{t=\tau_r}=\frac{\mu_0 c {\omega}\mu^2 N L w_0}{4\hbar}\mathrm{sin}\theta(\tau_r).
\end{equation}
$\theta(t)$ is a continuously differentiable function and 
\begin{equation}
    \frac{\mathrm{d}\theta}{\mathrm{d}t}=\frac{\mu_0 c {\omega}\mu^2 N L w_0}{4\hbar}\mathrm{sin}\theta(t)
    \label{dtheta_dt},
\end{equation}
with the initial time of $t\geq \tau_r$. The initial $\theta(\tau_r)$ is given by Eq.~(\ref{Bloch_angle}). The solution of Eq.~(\ref{dtheta_dt}) is 
\begin{equation}
    \theta(t)=2\mathrm{arctan}(e^{\frac{t-\tau_D}{\tau_W}}),
\end{equation}
where 
\begin{equation}
    \tau_W=\frac{4\hbar}{\mu_0 c {\omega}\mu^2 w_0 N L}
    \label{tau_W}
\end{equation}
is the characteristic duration of superradiance. $\tau_W$ can be expressed as
\begin{equation}
    \tau_W=\frac{16\pi \tau_{sp}}{3\lambda^2 w_0 N L},
\end{equation}
where $\tau_{sp}=\frac{3\pi \varepsilon_0 \hbar c^3}{\omega^3 \mu^2}$ is the spontaneous decay time. $\tau_D$ is determined by
\begin{equation}
    \theta(\tau_r)=\frac{\mu E_0}{\hbar}\int_0^{\tau_r} f(t) dt=2\mathrm{arctan}(e^{\frac{\tau_r-\tau_D}{\tau_W}}).
\end{equation}
$\tau_D$ is
\begin{equation}
    \tau_D=\tau_r-\tau_W \mathrm{ln}[\mathrm{tan}\frac{\theta(\tau_r)}{2}]=\tau_r- \frac{4\hbar \mathrm{ln}[\mathrm{tan}\frac{\theta(\tau_r)}{2}]}{\mu_0 c {\omega}\mu^2 w_0 N L}
    \label{tau_D solved}.
\end{equation}
The energy density of this two-level is written as
\begin{equation}
    E_N(t)=\frac{1}{2}\hbar \omega w_0 N \mathrm{cos}\theta({t})=-\frac{1}{2}\hbar \omega w_0 N \mathrm{tanh}(\frac{t-\tau_D}{\tau_W}).
    \label{E_t}
\end{equation}
The power $P_s$ of emission of superradiance per unit volume is
\begin{equation}
    P_s=-\frac{\mathrm{d}E_N(t)}{\mathrm{d}t}=\frac{\mu_0 c {\omega}^2\mu^2 {w_0}^2 N^2 L}{8}\mathrm{sech}^2(\frac{t-\tau_D}{\tau_W}).
\end{equation}
The electric field $\bar{E}_s(t)$ is
\begin{equation}
    \bar{E}_s(t,L)=\frac{\mu_0\omega c \mu w_0 N L}{2}\mathrm{sech}(\frac{t-\tau_D}{\tau_W}).
\end{equation}
The intensity $I_s$ (energy flux density $S$) of superradiance is 
\begin{equation}
    I_s=\frac{1}{2}\sqrt{\frac{\varepsilon_0}{\mu_0}}{\bar{E}_s}^2(t,L)=\frac{\mu_0 c {\omega}^2\mu^2 {w_0}^2 N^2 L^2}{8}\mathrm{sech}^2(\frac{t-\tau_D}{\tau_W})
    \label{I_s}.
\end{equation}

\section{The fitting of the seed pulse by the Gaussian profile}
\begin{figure}[ht]
\centering
\includegraphics[width=4in]{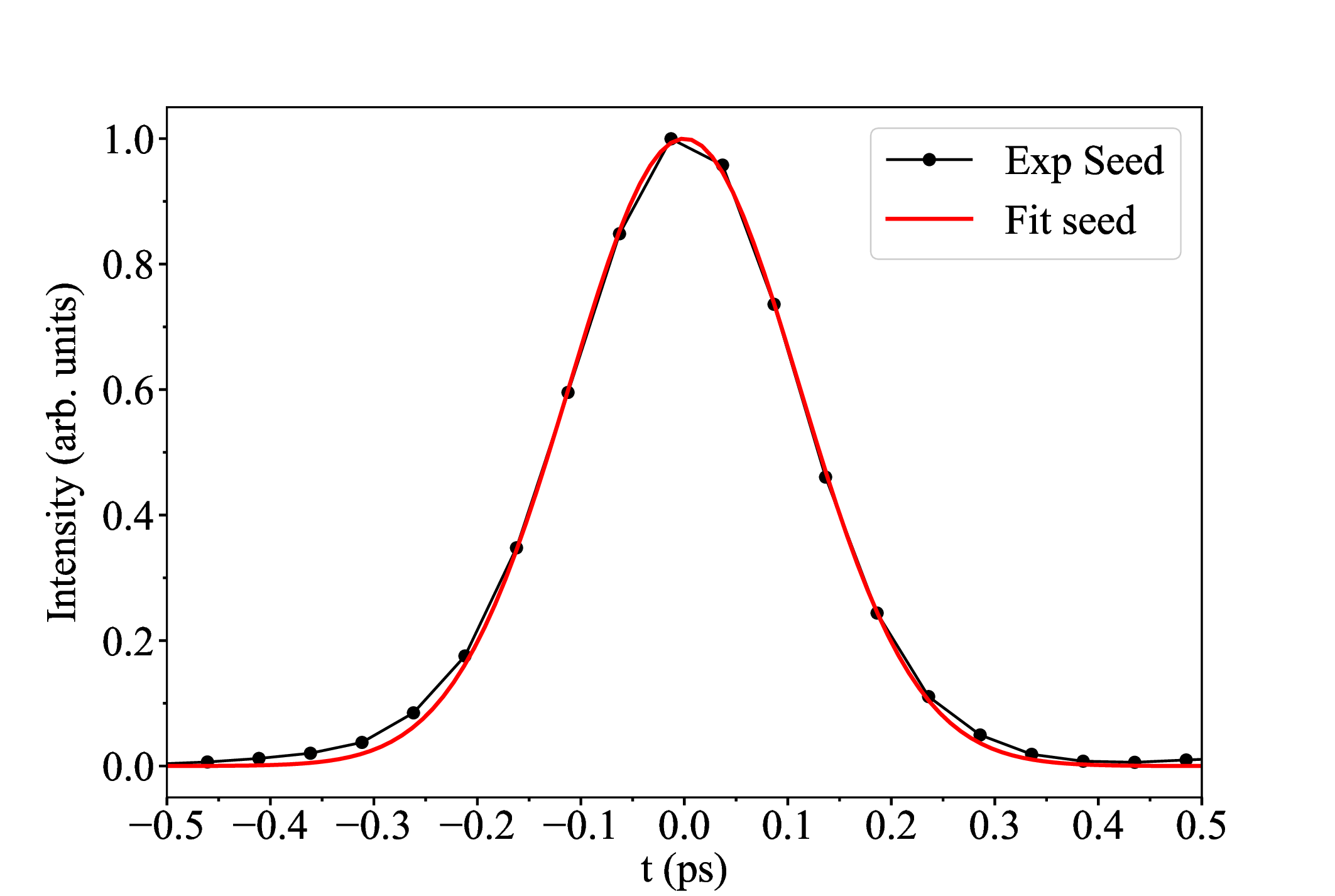}
\caption{The seed pulse of experimental data and fitting curve used in the calculation of the 391-nm superradiance as a function of gas pressure.}
\label{seed}
\end{figure}

In the experiment of measuring the 391-nm superradiance at different gas pressures, the seed pulse is shown in Fig.~\ref{seed}. It is fitted by the Gaussian pulse shape $e^{-4\mathrm{ln}2(\frac{t-\tau_s}{\tau_s})^2}$, with the FWHM $\tau_s=260$ fs. Because the fitting curve is in good agreement with the seed pulse, the Bloch angle $\theta(\tau_r)$ is obtained by integrating the Gaussian profile. It is noted that figure~\ref{seed} is the intensity envelope, so the envelope of electric field of the seed pulse is $e^{-2\mathrm{ln}2(\frac{t-\tau_s}{\tau_s})^2}$.